\documentstyle[11pt]{article}
\def\hybrid{\topmargin 0pt      \oddsidemargin 0pt
        \headheight 0pt \headsep 0pt
        \textheight 9in         
        \textwidth 6.25in }      
\catcode`\@=11
\def\marginnote#1{}
\newcount\hour
\newcount\minute
\newtoks\amorpm
\hour=\time\divide\hour by60
\minute=\time{\multiply\hour by60 \global\advance\minute by-\hour}
\edef\standardtime{{\ifnum\hour<12 \global\amorpm={am}%
        \else\global\amorpm={pm}\advance\hour by-12 \fi
        \ifnum\hour=0 \hour=12 \fi
        \number\hour:\ifnum\minute<10 0\fi\number\minute\the\amorpm}}
\edef\militarytime{\number\hour:\ifnum\minute<10 0\fi\number\minute}
\def\draftlabel#1{{\@bsphack\if@filesw {\let\thepage\relax
   \xdef\@gtempa{\write\@auxout{\string
      \newlabel{#1}{{\@currentlabel}{\thepage}}}}}\@gtempa
   \if@nobreak \ifvmode\nobreak\fi\fi\fi\@esphack}
        \gdef\@eqnlabel{#1}}
\def\@eqnlabel{}
\def\@vacuum{}
\def\draftmarginnote#1{\marginpar{\raggedright\scriptsize\tt#1}}
\def\draft{\oddsidemargin -.5truein
        \def\@oddfoot{\sl preliminary draft \hfil
        \rm\thepage\hfil\sl\today\quad\militarytime}
        \let\@evenfoot\@oddfoot \overfullrule 3pt
        \let\label=\draftlabel
        \let\marginnote=\draftmarginnote
   \def\@eqnnum{(\theequation)\rlap{\kern\marginparsep\tt\@eqnlabel}%
\global\let\@eqnlabel\@vacuum}  }

\catcode`@=12
\relax

\def\beq{\begin{equation}}
\def\eeq{\end{equation}}
\def\bea{\begin{eqnarray}}
\def\eea{\end{eqnarray}}
\def\nn{\nonumber}

\relax
\hybrid
 
\begin{document}

\begin{titlepage}
\begin{center}
February~1996 \hfill    PAR--LPTHE 96/06 \\
\hfill cond-mat/9602116\\[.4in]
{\large \bf {Effect of randomness in many coupled Potts models.}}\\[.3in]
	{\bf Pierre Pujol} \\
	{\it LPTHE\/}\footnote{Laboratoire associ\'e No. 280 au CNRS}\\
       \it  Universit\'e Pierre et Marie Curie, PARIS VI\\
       \it Universit\'e Denis Diderot, PARIS VII\\
	Boite 126, Tour 16, 1$^{\it er}$ \'etage \\
	4 place Jussieu\\
	F-75252 Paris CEDEX 05, FRANCE\\
\end{center}
\vskip .1in
\centerline{\bf ABSTRACT}
\begin{quotation}
Using 2-loop renormalisation group calculations, we study a system of
$N$ two-dimensional Potts models with random bonds coupled together by
their local energy density. This model can be seen as a generalization of
the random Ashkin-Teller model. We found that, depending on the sign of the
coupling term, the universality class of the system in the presence of
randomness is different. Under particular consideration, this model
presents an example of a first order phase transition rounded by
randomness. 

\end{quotation}

\end{titlepage}
\newpage

The effect of quenched randomness in two-dimensional systems having a
continuous phase transition in its absence has been studied for a large
variety of cases. In particular, using a renormalisation group (R.G.)
approach, the effect of uncorrelated impurities has been studied for the
two-dimensional Ising model \cite{dotising}, Baxter model \cite{dotdot} and
for the Potts model 
\cite{ludwig,dpp}. For most of these cases, the Harris criterion
\cite{harris} provide us a good method to see if disorder will change or
not the universality class. For first order phase transition systems,
the effect of randomness can be stronger. It was argued in \cite{imry} and
established in a rigorous way in \cite{aizenman} that for two-dimensional
systems the transition becomes continuous in the presence of
impurities. Recently, it was shown by Cardy \cite{cardy} that such phenomenon
effectively happens for a large class of systems presenting a fluctuation
driven first order transition in the pure case. 
In view of the results obtained there, it was suggested that the
Ising-like transition could be a general feature of two-dimensional systems
with randomness. 
Of particular interest is
the random $N$ color Ashkin-Teller model considered in this work. This
model can be easily generalized to $N$ coupled $q$-state Potts model (with
$2\leq q \leq 4$) with random bonds. The aim of this paper is to compare
the behavior of this model in the pure case and in the presence of quenched
randomness for generic $N$. Our result is that when impurities are
added, the system will present a continuous
transition but with different universality classes depending on the sign
of the coupling between the models. In any case, the transition is not
Ising-like.

Our model consists of $N$ two-dimensional $q$-state Potts models near the
critical point 
coupled together by their energy operators. This model can be considered as
a generalization of the $N$ color Ashkin-Teller model with $2 \leq q <
4$. The Hamiltonian of the system has the following form:
\beq
\label{nonrandham}
H = \displaystyle\sum_{i=1}^N H_{0,i} +\int d^2x ~ m \displaystyle\sum_{i=1}^N 
\varepsilon_{i}(x) - g \int d^2x \displaystyle\sum_{i \neq j}^N \varepsilon_{i}
\varepsilon_{j}(x)~.
\eeq
$H_{0,i}$ are the Hamiltonians of the $N$ Potts models at the
critical temperature, $m$ is the reduced temperature, $\varepsilon$
corresponds to the energy operator of the pure models 
and the last term in (\ref{nonrandham}) is the
coupling between the different models. This model was well studied
in the case $q=2$ which is the $N$ color Ashkin-Teller model. For $N=2$ it
turns to be integrable \cite{baxkadwu}. 
For $N>2$ the large scale behavior
depends on the sign of $g$: For $g<0$ it presents an Ising type second order
phase transition while for $g>0$ a fluctuation driven first order phase
transition is present \cite{fradkin,shankar}. Recently, it was shown
\cite{vays} that for $N=2$, $q>2$ the model
is still integrable presenting now a mass generation indicating again a
first order phase transition. The situation is not so clear for $N>2$. In
the lack of an exact solution we can give a R.G. approach by studying the
large scale behavior of the system for different values of $g$. This was
done in \cite{ludwig, dpp} for the case $N \rightarrow 0$ in the context of
the Potts model with random bonds. A non trivial fixed point was found
with critical exponents for the specific heat and magnetic susceptibility
which differ from the ones of the pure model.

For generic $N$, 
in the limit $m \rightarrow 0$, the
2-loop R.G. equations of (\ref{nonrandham}) can be obtained from \cite{dpp}:
$$
\beta(g) \equiv {dg\over dln(r)} = \epsilon g(r) + 4 \pi (N-2) g^2(r) -16
\pi^2 (N-2) g^3(r) + O(g^4(r))
$$
$$
\gamma_{\varepsilon} \equiv
{d ln(Z_\varepsilon(r)) \over d ln(r)} =   4\pi (N-1) g
 -   8 \pi^2 (N-1) g^2 + O(g^3)
$$
\bea
\label{rg1}
\gamma_{\sigma} \equiv
{dln(Z_\sigma(r))\over dln(r)} =  (N-1)g^2(r) \pi^2 \epsilon
\left[1 + 2
{\Gamma^2(-{2\over3})\Gamma^2({1\over6})\over
\Gamma^2(-{1\over3})\Gamma^2(-{1\over6})}\right] \nn \\
+ 4 (N-1)(N-2) \pi^3 g^3(r)+ O(g^4)
\eea
Here, $Z_\varepsilon$ and $Z_\sigma$ are respectively the renormalisation
constants of the energy and spin operators and $\epsilon = 2 - 2
\Delta_{\varepsilon}$ where $\Delta_{\varepsilon}$ is the physical
dimension of the energy operator of the pure model ($\epsilon$
 differs from that of
\cite{dpp} by a factor of $-3$, so $\epsilon = 0$ for $q=2$ and 
$\epsilon = {2\over 5}$ for $q=3$). 
For the case $N>2$, we can see that if
$g>0$ the coupling constant flow far from our perturbative region. Even if
a definite proof is not given, a comparison with the case $q=2$ seems to
tell us that a mass gap is dynamically generated indicating a first order
phase transition. The situation is completely different for $g<0$. In this
case, we can see that there is a non trivial infrared fixed point at:
\beq
\label{fixed1}
g_c = -{\epsilon \over 4
\pi (N-2)} + {\epsilon^2 \over 4\pi (N-2)^2} + O(\epsilon^3)
\eeq
The critical exponents associated with the energy and spin
operators in this new infrared fixed point are given by:
$$
\Delta'_{\varepsilon} = \Delta_{\varepsilon} - \gamma_\varepsilon(g_c)
$$
\beq
\label{expe1}
= \Delta_{\varepsilon} + {(N-1)\over (N-2)} \epsilon
-  {(N-1)\over 2(N-2)^2} \epsilon^2 + O(\epsilon^3)
\eeq
and
$$
\Delta'_{\sigma} = \Delta_{\sigma} - \gamma_{\sigma}(g_c)
$$
\beq
\label{expsig1}
=  \Delta_{\sigma} - \left( {(N-1)\over 8 (N-2)^2}\right)
{\Gamma^2(-{2\over3})\Gamma^2({1\over6})\over\Gamma^2(-{1\over3})
\Gamma^2(-{1\over6})} \epsilon^3 + O(\epsilon^4)
\eeq
The results obtained in \cite{ludwig, dpp} for the case of the quenched
random case can be reproduced by just putting $N=0$ in the
formula above and keeping in mind that these results then make sense
only for $g>0$, which in this case $g$ corresponds to the variance of the
probability distribution for the random bonds.

Let's consider now adding quenched randomness which couples to the energy
operator (e.g. the local energy density). This can be done by introducing
in (\ref{nonrandham}) a position dependent random mass term $m \rightarrow
m(x)$  where $\overline{m(x)} = 0$ and $\overline{m(x)m(y)} = \Delta \delta
(x-y)$, with $\Delta > 0$. 
Using the replica method, we can introduce $n$ copies of the system and
average over a Gaussian distribution for $m$, this will give us a
Hamiltonian of $nN$ Potts models coupled by their energy operators. The
replicated Hamiltonian is:
\beq
\label{randham}
H = \displaystyle\sum_{i,a} H_{0,i}^a 
- g \int d^2x \displaystyle\sum_{i \neq j,a} \varepsilon_{i}^a
\varepsilon_{j}^{a}(x)
- \Delta \int d^2x \displaystyle\sum_{<i,j,a,b>} \varepsilon_{i}^a
\varepsilon_{j}^b (x)
\eeq
where indices $i$, $j$ runs from $1$ to $N$ and $a$, $b$ from $1$ to $n$ 
and $<\cdots >$ means that terms with same replica indices are omitted. 
As explained in \cite{ludwig}, this is because these terms produce irrelevant
operators or trivial contributions which could just give a shift in the
critical temperature. 
This model was studied at 1-loop level in \cite{cardy} for the case $q=2$ (the
random Ashkin-Teller model). It was shown that when randomness is present,
the R.G. trajectories for the coupling constants run away in a first stage
from the U.V. fixed point but finally curl around and approach the fixed
point corresponding to $N$ decoupled Ising models.

The procedure to obtain the 2-loops R.G.  equations for (\ref{randham}) is
the same as in \cite{dpp}. After some combinatory, we find, taking
directly the limit $n \rightarrow 0$:
$$
\dot{g} = \epsilon g + (N-2) g^2 - 2g\Delta - (N-2) g^3 - (2N-5)g^2 \Delta
+ 4g \Delta^2
$$
\beq
\label{rg2}
\dot{\Delta} = \epsilon \Delta -2 \Delta^2 + 2(N-1) \Delta g +2 \Delta^3
- (N-1) g^2 \Delta - 2(N-1)\Delta^2 g
\eeq
where for simplicity we have made the change $g\rightarrow 4\pi g$, $\Delta
\rightarrow 4\pi \Delta$.
The first step in studying the flow given by (\ref{rg2}) is to look for the
fixed points $\dot{g} =\dot{\Delta} =0$. Let's first consider the case
$N>2$. Then, we find four fixed points given by:
\beq
\label{qfixed1}
g=0~~~;~~~\Delta =0
\eeq
\beq
\label{qfixed2}
g=-{\epsilon \over (N-2)} + {\epsilon^2 \over (N-2)^2} +
O(\epsilon^3)~~~;~~~\Delta =0 
\eeq
\beq
\label{qfixed3}
g=0~~~;~~~\Delta = {\epsilon \over 2} + {\epsilon^2 \over 4}
+ O(\epsilon^3)
\eeq
\beq
\label{qfixed4}
g={\epsilon^2 \over 2N} + O(\epsilon^3)~~~;
~~~\Delta = {\epsilon \over 2} + {(3N-2)\epsilon^2 \over 4N} 
+ O(\epsilon^3)
\eeq
The others fixed points present in (\ref{rg2}) are of order of constant
and so beyond our perturbative region. The non trivial fixed points 
(\ref{qfixed2}) and (\ref{qfixed3}) are the ones present in the case of
$N$ non random coupled models and one random model respectively. There is
however a new fixed point (\ref{qfixed4}) that can not be seen at the
1-loop level. As usual, we can study the stability of each of these fixed
points by re-expressing (\ref{rg2}) around the solutions above
$g=g_c+\delta g$, $\Delta = \Delta_c + \delta \Delta$ and keeping only the
smallest order in $\epsilon$. This will give us a linear system:
$$
\left( {\delta \dot{g} \atop \delta  \dot{\Delta}}\right) = A
\left( {\delta g \atop \delta  \Delta}\right)  
$$
The information about the stability of each of these fixed points can be
obtained by calculating the eigenvalues of $A$ in each of these cases. In
this way it is easy to see that (\ref{qfixed1}) is unstable
(for $\epsilon >0$),
(\ref{qfixed2}) and (\ref{qfixed4}) are both stable while (\ref{qfixed3})
is stable only when $g=0$. This result tell us that the random fixed point
found in \cite{ludwig,dpp} can not be reached if a small coupling between
the different Potts model is added. To study the coupling constant flow 
, let's first suppose that the initial conditions are $\Delta (0)~,~ 
g(0) > 0$.
It is useful in a first stage to keep just the first and
second order terms in (\ref{rg2}).  
In this case, it can be shown that for
generic $\epsilon$ the solution is given by:
\beq
\Delta (l) = \epsilon \tau \Delta_{0}(\tau) ~~~;~~~
g(l) = \epsilon \tau g_{0}(\tau)
\eeq
where $\tau = e^{\epsilon l} / \epsilon $, $l$ is the R.G. scale 
parameter and $\Delta_0$, $g_0$ are the solutions of (\ref{rg2}) with
$\epsilon = 0$. 
A closed form for $\Delta_0$ and $g_0$ has been given
in \cite{cardy}, it is easy to see that for $\tau \rightarrow \infty$ we
have $g_0 (\tau) << \Delta_0 (\tau)$ and $\Delta_0 (\tau) \sim {1\over 
2 \tau}$. So, the trajectories of the coupling
constants for generic $\epsilon$ will flow toward the region: 
$g(l) << \Delta (l)$ and $\Delta (l) \sim {\epsilon \over 2}$. This is
precisely the region where third order terms in (\ref{rg2}) become
important indicating that trajectories will asymptotically reach
the point (\ref{qfixed4}). 
We can now calculate the critical exponents associated with the energy
and spin operator. Since the difference between the points (\ref{qfixed3})
and (\ref{qfixed4}) is of order $\epsilon^2$ we can not see at this order
in perturbation any difference in the spin critical exponent
compared to that found in \cite{dpp} for one Potts model with
random bonds. For the energy operator, the correction to the critical 
exponent up to second order is given by:
$$
\Delta_{\varepsilon}' = \Delta_{\varepsilon} + \Delta_c - (N-1)g_c
- {\Delta_{c}^2 \over 2} + \cdots
$$
\beq
= \Delta_{\varepsilon} + {\epsilon \over 2} + {\epsilon^2 \over 8}
+ O(\epsilon^3)
\eeq
giving surprisingly for generic $N$ the same result as for $N=1$.
Repeating the same arguments, we can show
that for initial conditions $\Delta (0)> 0~,~g(0)<0$ the R.G. 
trajectories will now flow toward the point (\ref{qfixed2}) which is
the nonrandom fixed point. The critical exponents will be given
by (\ref{expe1}) and (\ref{expsig1}) indicating that for this sign of
$g$ randomness will not change the large scale behavior of the system.

Let's finally turn to the particular case $N=2$. As was said before,
the non random case is exactly integrable \cite{vays} giving
a mass for any non zero value of $g$. It is interesting to see if
randomness will increase the order of the transition or will
keep a non zero mass in the theory. The R.G. equations are simply
obtained by rewriting (\ref{rg2}) for $N=2$:
$$
\dot{g} = \epsilon g - 2g\Delta + g^2 \Delta
+ 4g \Delta^2
$$
\beq
\label{rg3}
\dot{\Delta} = \epsilon \Delta -2 \Delta^2 + 2 \Delta g +2 \Delta^3
-  g^2 \Delta - 2\Delta^2 g
\eeq
The fixed points (\ref{qfixed1}) and (\ref{qfixed3}) are still
presents with the same type of stability and (\ref{qfixed4})
becomes now:
\beq
\label{qfixed5}
g={\epsilon^2\over 4}~~~;~~~\Delta=
{\epsilon + \epsilon^2\over 2}
\eeq
But now, as expected, (\ref{qfixed2}) has disappeared. In the region
$g>0$ the arguments given for generic $N$ are still valid and 
trajectories will flow toward the point (\ref{qfixed5}) giving at
this order the same spin and energy 
exponents as for generic $N$.
The difference appear for initial conditions with $g<0$. In this
case, the R.G. trajectories will flow far from our perturbative
region. With the results above we can not say if trajectories will flow
toward a massive theory or a non perturbative fixed
point. A naive analysis of (\ref{rg2}) in this case indicates that
trajectories will reach asymptotically the integrable line $\Delta =0$
which gives a massive theory. This would be in contradiction with
Imry and Wortis arguments \cite{imry} and statements given in \cite{aizenman}. 
A similar situation happens in the case $q=2$, 
$N=2$ (the random Baxter model) studied in \cite{dotdot}. This example of a
system with quenched randomness violate the Harris criterion which is in
some sense the second order transition version of Imry-Wortis argument.

In this paper we have analyzed the behavior in presence of 
quenched randomness of
a model which present a very rich structure. Depending on the sign of the
coupling $g$, the nature of the transition is different in the
pure case. We found that in the random case, for $N>2$,
whatever the sign of $g$ is, the transition is continuous 
and different from that of the random Ising model, but the
universality classes are different for $g>0$ and
$g<0$. Moreover, in the case $g<0$ the disorder doesn't change the
universality class of the model. This apparent contradiction of Harris
criterion is due to the fact that randomness, which is relevant at the tree
level, is driven to be irrelevant by the coupling term. The behavior of the
model in the region $g>0$ is also surprising: the fixed point of $N>2$
coupled Potts models in the presence of impurities is not the same as found in
\cite{ludwig,dpp} in the case $N=1$. The spin and energy exponents remains
however, at this order in perturbation, the same for any value of $N$. 

The
particular case $N=2$ is more complicated; for $g>0$ the behavior of
the R.G. flow is similar to the one of the case 
$N>2$. Since the pure case is integrable
and is a massive theory, we have a system which certainly gives a first
order phase transition in the pure case. The presence of randomness makes
however the transition continuous.
This give a new example of a first
order transition rounded by randomness. It is not clear however what finally
happens for $g<0$ and only a non-perturbative treatment would give a
definitive answer.

I would like to thank Vl. S. Dotsenko and M. Picco for many helpful
suggestions and conversations.

\end{document}